\documentclass[preprint,showpacs,preprintnumbers,eqsecnum,
amsmath,amssymb,pra]{revtex4}
\usepackage{epsfig}

\begin{document}

\title{Exact results for state-to-state transition probabilities

in the multistate Landau-Zener model

by non-stationary perturbation theory
}

\author{M.~V.~Volkov$^{1,2}$}
\email[E-mail: ]{mvvolkov@mail.ru}
\author{V.~N.~Ostrovsky$^1$}
\affiliation{$^1$ V.~Fock Institute of Physics, St Petersburg
State University, St Petersburg 198504, Russia}
\affiliation{$^2$ Department of Physics, AlbaNova University Center,
Stockholm University, 106 91 Stockholm, Sweden }

\begin{abstract}

Multistate generalizations of Landau-Zener model are studied by
summing entire series of perturbation theory. A new technique for
analysis of the series is developed. Analytical expressions for
probabilities of survival at the diabatic potential curves with
extreme slope are proved. Degenerate situations are considered
when there are several potential curves with extreme slope. New
expressions for some state-to-state transition probabilities are
derived in degenerate cases.

\end{abstract}


\maketitle

\section{Introduction}
The famous Landau-Zener two-state model, introduced and solved in
1932 by Landau \cite{Land}, Zener \cite{Zen}, Majorana \cite{M}
and St\"{u}ckelberg \cite{St} finds many applications in atomic
physics and beyond. This is due to its virtue of describing
generic case of non-adiabatic transitions in quantum mechanics.
The main feature of the exactly solvable quantum model is the
linear dependence of the matrix Hamiltonian on time. The model
allows the derivation of exact expression for the state-to-state
transition probability.

The natural generalization of the two-state model is the model
with arbitrary (but still finite) number of states, $N$. The
linear dependence of matrix Hamiltonian on time is retained ${\bf
H}(t)={\bf A}+{\bf B}t$, where ${\bf A}$ and ${\bf B}$ are
time-independent $N \times N$ matrices. Hereafter we show by bold
type the operators and vectors in $N$-dimensional linear space.
The lower case characters are used to denote vectors in this
space while the capital characters denote matrix operators.

Without loss of generality one might assume that the basis is
chosen in such a way that the Hermitian matrix ${\bf B}$ is
diagonal, $B_{jk}=\beta_j\delta_{jk}$, where the parameters
$\beta$ have the meaning of slopes of linear diabatic potential
curves. The so chosen basis is known as the {\it diabatic basis}.
The non-diagonal elements of matrix ${\bf A}$ have the meaning of
couplings between the diabatic states, $A_{jk}=V_{jk}$. The
diagonal matrix elements of ${\bf A}$ play a different role. It
is convenient to introduce for them a special notation,
$\varepsilon_j=A_{jj}$ (these notations are the same as in our
preceding studies \cite{SurPr,No-go,reply}). The diagonal matrix
elements of the Hamiltonian ${\bf H}(t)$ are referred to as {\it
diabatic potential curves}. In the case of the multistate
Landau-Zener model, they are linear in time, $E^{\rm
dia}_j(t)=\beta_j t + \varepsilon_j$.

The problem is to solve the non-stationary Schr{\" o}dinger
equation
\begin{eqnarray}\label{SE}
i\frac{d{\bf c}}{dt}={\bf H}(t)~{\bf c}~,
\end{eqnarray}
and to find S-matrix. Generally speaking the full exact solution
of (\ref{SE}) is not available. The known exact solutions
\cite{Dem-Osh,Ostr-Nak,DOgen1,DOgen2} refer to special choices of
the model parameters $\beta_j$, $\varepsilon_j$, $V_{jk}$, such
that the quantum interference oscillations do not appear in the
transition probabilities. Furthermore, even in the case of the
most general form of matrix Hamiltonian one can {\it exactly}\/
find two elements of S-matrix which correspond to survival on the
diabatic potential curves with extremal (maximum or minimum)
slopes. The simple formula for such elements was originally
guessed by S.~Brundobler and V.~Elzer \cite{BE} based on
numerical calculations. The proof of Brundobler-Elzer (BE)
formula was carried out recently by several different ways.
A.~V.~Shytov obtained this formula via treatment within the
contour integration approach \cite{Shyt}. M.~V.~Volkov and V N
Ostrovsky carried out the proof using non-stationary perturbation
theory \cite{SurPr}. However there are some oversights in this
proof, as B.~E.~Dobrescu and N.~A.~Sinitsyn indicated in the
comment to this paper \cite{Comment}. The comment contains a new
proof of BE formula partly based on developments by Volkov and
Ostrovsky; at the crucial step it essentially uses results for
the bow-tie model \cite{Ostr-Nak} exactly solved by Ostrovsky and
Nakamura.

The objective of the present study is to provide a proof of the BE
formula which is devoid of deficiency of the previously suggested
proof being fully based on analysis of non-stationary
perturbation theory and summation of an entire perturbative
expansion. Compared to the case of a Hamiltonian ${\bf H}_{\rm
bound}(t)$ with all the matrix element bounded [$\left(H_{\rm
bound}\right)_{jk}(t) < a$ for all times $t$] the case of the
multistate Landau-Zener Hamiltonian provides important specifics.
The emerging integrals typically contain highly oscillating
exponential factors that ensure integral convergence. For some
choice of parameters in the integrand the oscillations vanish
which means that the integral is a singular function of
parameters. These singularities are to be treated in the analysis
with proper care; albeit namely the presence of singularities
allows a closed-form evaluation for each term of the entire
series with subsequent analytical summation.

In the main Section \ref{proofBE} we develop a new approach to
treat the singularities. The preliminary Section \ref{nonstpert}
introduces notations and contains a general description of the
perturbative series. In distinction to the scheme suggested by
Dobrescu and Sinitsyn \cite{Comment}, our proof (Section
\ref{proofBE}) does not use results of any exactly solvable model.
We believe that such a complete treatment of the perturbative
expansion with analytical summation of series is of general
interest.

Another goal of our study is to consider some degenerate cases
(Section \ref{degen}). Here more state-to-state probabilities can
be evaluated, up to the fully degenerate multistate model where an
entire matrix of state-to-state transition probabilities is found
(Section \ref{fully}).

\section{Non-stationary perturbation theory} \label{nonstpert}

The well-known formula for transition probability for two-state
linear model was derived by Zener by reducing the Schr{\" o}dinger
equation to an equation for a hyperbolic cylinder function
\cite{Zen}. Majorana \cite{M} used contour integration method in a
complex plane to solve the same problem. Much later Kayanuma
suggested an alternative approach \cite{Kay,Kay1} where the
non-stationary perturbation theory is used. As discussed in the
Introduction, we in the present paper provide a generalization of
this method to the multistate case.

The non-stationary Schr{\" o}dinger equation (\ref{SE}) might be
written as the set of N coupled first-order differential
equations:
\begin{eqnarray} \label{SE1}
i \frac{dc_j}{dt} = \varepsilon_j c_j + \beta_j t c_j + \sum_{k
\neq j} V_{jk} c_k ~, \qquad j, \: k = 1, 2, \ldots N ~.
\end{eqnarray}
After phase transformation which eliminates the diagonal elements
on the right-hand side of equations (\ref{SE1}) it takes the form
\begin{eqnarray} \label{SE2}
 i \frac{da_j}{dt} = \sum_{k \neq j} V_{jk} \exp\left[ i \left(
(\varepsilon_j-\varepsilon_k)t + \frac{1}{2} (\beta_j-\beta_k) t^2
\right) \right] a_k ~, \qquad j, \: k = 1, 2, \ldots N ~.
\end{eqnarray}
The integral form of this equation
\begin{equation} \label{INTSE}
 a_j(t)= a_j(-\infty) -i\int_{-\infty}^{t} \, dt_0 \,
\sum\limits_{k\neq j}V_{jk}
\exp\left[i(\varepsilon_j-\varepsilon_k)t_0+ \frac{i}{2}
(\beta_j-\beta_k) t_0^2\right] a_k(t_0)
\end{equation}
is convenient for an iterative solution. The successive
approximations, $a_j^{(n)}(t_{n})$, are found by iterations:
\begin{eqnarray} \label{ITSE}
 a_j^{(n+1)}(t_{n+1})=a_j^{(0)}(-\infty)
-i\int_{-\infty}^{t_{n+1}} dt_n \sum\limits_{k\neq j}V_{jk}
\exp\left[i(\varepsilon_j-\varepsilon_k)t_n+
i\frac{1}{2}(\beta_j-\beta_k)t_n^2\right]a_k^{(n)}(t_n)~.
\end{eqnarray}
We use label 1 for the initially populated state, so that initial
populations $a_j(-\infty)$ are
\begin{equation} \label{initial}
a_j(-\infty)=\delta_{j1} ~.
\end{equation}
Then transition probability to $j$-th state is
\begin{eqnarray} \label{PR}
P_{1j}= \left|\lim_{n \rightarrow \infty}
 \, a_j^{(n)}(+\infty) \right|^2 ~.
\end{eqnarray}
In the next formula we introduce a vector-function of time ${\bf
f}(t) =\{f_1(t), f_2(t)\ldots f_N(t)\}$, which is a vector in
$N$-dimensional linear space. The operator ${\bf T}$ is $N\times
N$ matrix; it transforms the vector-function ${\bf f}(t)$ into
another vector-function with components:
\begin{eqnarray}\label{opT}
{\left[\hat{\bf T}\bf f\right]}_j (t_{n+1}) & \equiv & (-i)
\sum\limits_{{k=1}\atop{k \neq j}}^{N}
V_{jk}\int_{-\infty}^{t_{n+1}} dt_n
\exp\left[i(\varepsilon_j-\varepsilon_k) t_n
+ \frac{i}{2} (\beta_j-\beta_k) {t_n}^2\right]f_k(t_n) ~.
\end{eqnarray}
With respect to the time variable the operator ${\bf T}$ is an
integral operator. Our equations (\ref{ITSE}) can be written as
\begin{eqnarray}\label{ITSE1}
{\bf a}^{(n+1)}={\bf a}^{(0)}+{\hat{\bf T}}{\bf a}^{(n)}~,
\end{eqnarray}
where dependence on time is implicit. The zero iteration ${\bf a
}^{(0)}$ is defined by the initial conditions (\ref{initial}):
$a^{(0)}_j = \delta_{j1}$.

We further introduce the vector ${\bf d}^1$ in $N$-dimensional
linear space by a formula describing its components $d^1_j$:
\begin{eqnarray} \label{opd1}
d^{(1)}_j(t)\equiv -iV_{j1}\int_{-\infty}^{t}dt_1
\exp\left[i(\varepsilon_j-\varepsilon_1) t_1
+ \frac{i}{2}(\beta_j-\beta_1)t^2_1\right] ,
\quad
j \neq 1 ~.
\end{eqnarray}
The $j=1$ component $d^{(1)}_1$ is assumed to be zero by
definition. Similarly, the vector ${\bf d}^{(m)}_j$ $(m \geq 2)$
is given as
\begin{eqnarray} \label{opdm}
d^{(m)}_j (t) & \equiv & {(-i)}^m \sum\limits_{{k_{m-1}\neq
j}}^NV_{jk_{m-1}}\sum\limits_{{k_{m-2}\neq
k_{m-1}}}^NV_{k_{m-1}k_{m-2}}\ldots\sum\limits_{{k_2\neq
k_3}}^NV_{k_3k_2} \sum\limits_{{k_1\neq k_2}\atop{k_1\neq 1}}^N
V_{k_2k_1}V_{k_11}
\nonumber \\ && \times
\int_{-\infty}^{t}dt_m\int_{-\infty}^{t_{m}}dt_{m-1} \ldots
\int_{-\infty}^{t_{2}} dt_{1} \,
\nonumber \\ && \times
\exp\left[i(\varepsilon_j-\varepsilon_{k_{m-1}})t_m +
i\sum\limits_{i=2}^{m-1}(\varepsilon_{k_i}-\varepsilon_{k_{i-1}})
t_{i}+i(\varepsilon_{k_1}-\varepsilon_1)t_{1}\right]
\nonumber \\ && \times
\exp\left[\frac{i}{2}(\beta_j-\beta_{k_{m-1}}) t_m^2 + \frac{i}{2}
\sum\limits_{i=2}^{m-1}(\beta_{k_i}-\beta_{k_{i-1}}) t^2_{i}
+\frac{i}{2}(\beta_{k_1}-\beta_1) t^2_{1} \right] ~.
\end{eqnarray}
If the couplings are small, then the order of magnitude estimates
are $T \sim V$, ${{\bf d}}^{(m)} \sim V^m$. Note the important
relations between operator ${\bf \hat{T}}$ and vectors ${\bf
d}^{(m)}$:
\begin{eqnarray}
\hat{\bf T}{\bf d}^{(m)} &=&{\bf d}^{(m+1)}~,\qquad
m=1, \, 2, \, \ldots ~,
\nonumber\\
\hat{\bf T}{\bf a}^{(0)}&=&{\bf d}^{(1)}~.
\end{eqnarray}
Using these relations and equation (\ref{ITSE1}) we express
the $n$-th iteration to ${\bf a}$ as
\begin{eqnarray}\label{ITSE2}
{\bf a}^{(n)}={\bf a}^{(0)}+\sum\limits_{m=1}^{n}{{\bf d}}^{(m)}~.
\end{eqnarray}
Formula (\ref{ITSE2}) is the basis for all subsequent analysis.
In order to find some transition amplitude one should evaluate
the corrections (\ref{opdm}) to all orders $m$ in the limit $t
\rightarrow +\infty$, then sum up all corrections using equation
(\ref{ITSE2}) with $n \rightarrow +\infty$. The sought for
probability is given by formula (\ref{PR}).

\section{Proof of the Brundobler-Elzer formula} \label{proofBE}

\subsection{Preliminary transformations: change of variables}

Consider the case when the initially populated non-degenerated
diabatic potential curve has ES, i.e. its slope is the largest
$(\beta_1=\max_j \beta_j)$ or the smallest $(\beta_1=\min_j
\beta_j)$ of all slopes. Here we set out to find the survival probability
on such a potential curve. The general vector formula
(\ref{ITSE2}) for the first component reads
\begin{eqnarray}\label{ITSE21}
{a}_1^{(n)}={a}_1^{(0)}+\sum\limits_{m=1}^{n}d^{(m)}_1=
1+\sum\limits_{m=1}^{n}d^{(m)}_1~.
\end{eqnarray}
The arbitrary term in the sum is given by (\ref{opdm}) and
(\ref{opd1}). In the limit $t\rightarrow\infty$ and after reducing
the brackets we obtain:
\begin{eqnarray}\label{dm1}
d_1^{(m)}(\infty)&=&{(-i)}^m
\sum\limits_{k_{m-1}\neq1}^NV_{1k_{m-1}}\sum\limits_{k_{m-2}\neq
k_{m-1}}^NV_{k_{m-1}k_{m-2}}\ldots\sum\limits_{k_2\neq
k_3}^NV_{k_3k_2} \sum\limits_{{k_1\neq k_2}\atop{k_1\neq 1}}^N
V_{k_2k_1}V_{k_11}
\nonumber \\ && \times
\int_{-\infty}^{\infty}dt_m\int_{-\infty}^{t_{m}}dt_{m-1} \ldots
\int_{-\infty}^{t_{2}} dt_{1}
\nonumber \\ && \times
\exp\left[i\varepsilon_1(t_m-t_{1})+
i\sum\limits_{i=1}^{m-1}\varepsilon_{k_i}(t_{i}-t_{i+1})\right]
\nonumber \\ && \times
\exp\left[\frac{i}{2}\beta_1({t_m}^2-t^2_{1})+
\frac{i}{2}\sum\limits_{i=1}^{m-1}\beta_{k_i}
(t^2_{i}-t^2_{i+1})\right] .
\end{eqnarray}
Let us now introduce new integration variables
$\{x_{1},\ldots,x_m\}$ such that:
\begin{eqnarray} \label{xt}
x_m &=& t_m~,
\qquad x_m \in (-\infty, \, \infty);
\nonumber \\
x_j&=&t_{j+1}-t_j~,
\qquad
x_j\in (0, \, \infty),
\qquad
j=1, \, 2 \, \ldots, \, m-1~.
\end{eqnarray}
The important advantage of this transformation is that the ranges
of variation of the new variables are simple and unambiguous, cf.
discussion in Refs.~\cite{Comment,reply}. The Jacobian of the
transformation is equal to ${(-1)}^{m-1}$, the inverse
transformation is given by
\begin{eqnarray} \label{tx}
t_j&=&x_m-\sum\limits_{k=j}^{m-1}x_k~,\qquad
j=1, \, 2, \ldots, \, m-1~;
\nonumber\\
t_m&=&x_m~.
\end{eqnarray}
In order to express the integrand in (\ref{dm1}) in new variables
the following formulas are useful:
\begin{eqnarray}
t_m-t_{1}&=&\sum\limits_{k=1}^{m-1}x_k~;
\nonumber \\
t^2_m-t^2_{1} & = & 2x_m\sum\limits_{k=1}^{m-1}x_k-{\left(
\sum\limits_{k=1}^{m-1}x_k\right)}^2 ;
\nonumber \\
t^2_{i}-t^2_{i+1}&=&2x_m(-x_{i})+x_{i}
\left(x_{i}+ 2\sum\limits_{k=i+1}^{m-1}x_k \right)\qquad
i=1,\ldots,m-2~;\nonumber\\
t^2_{m-1}-t^2_{m}&=&-2x_mx_{m-1}+x^2_{m-1}~.
\end{eqnarray}
In new variables the integral is cast as
\begin{eqnarray}
d_1^m&=&{(-i)}^m \sum\limits_{{k_{m-1}\neq
1}}^NV_{1k_{m-1}}\sum\limits_{{k_{m-2}\neq
k_{m-1}}}^NV_{k_{m-1}k_{m-2}}\ldots\sum\limits_{{k_2\neq
k_3}}^NV_{k_3k_2} \sum\limits_{{k_1\neq k_2}\atop{k_1\neq 1}}^N
V_{k_2k_1}V_{k_11} \nonumber \\ &&
\times\int_{-\infty}^{\infty}dx_m\int_0^{\infty}dx_{m-1}\cdots
\int_0^{\infty}dx_{1}
\exp\left[i\varepsilon_1\sum\limits_{n=1}^{m-1}x_n-
i\sum\limits_{n=1}^{m-1}\varepsilon_{k_n}x_{n}\right]
\nonumber\\
&& \times \exp\left[\frac{i}{2}\beta_1\left(2x_m\sum\limits_{n=1}^{m-1}x_n-{\left(
\sum\limits_{n=1}^{m-1}x_n\right)}^2\right)\right]\nonumber
\\&&\times \exp\left[\frac{i}{2}\sum\limits_{n=1}^{m-2}\beta_{k_n}
\left(2x_m(-x_{n})+x_{n} \left(x_{n}+
2\sum\limits_{j=n+1}^{m-1}x_j \right)\right)\right]\nonumber\\
&&\times\exp\left[\left(-2x_mx_{m-1}+x^2_{m-1}\right)\beta_{k_{m-1}}\right]~.
\end{eqnarray}
The integration over $dx_m$ in infinite limits gives a
$\delta$-function. After reducing brackets in the exponents one
obtains
\begin{eqnarray}\label{dm1-1}
d_1^m&=&{(-i)}^m \sum\limits_{{k_{m-1}\neq
1}}^NV_{1k_{m-1}}\sum\limits_{{k_{m-2}\neq
k_{m-1}}}^NV_{k_{m-1}k_{m-2}}\ldots\sum\limits_{{k_2\neq
k_3}}^NV_{k_3k_2} \sum\limits_{{k_1\neq k_2}\atop{k_1\neq 1}}^N
V_{k_2k_1}V_{k_11} \nonumber \\ &&
\times\int_0^{\infty}dx_{m-1}\cdots \int_0^{\infty}dx_{1}
\exp\left[i\sum\limits_{n=1}^{m-1}(\varepsilon_1-\varepsilon_{k_n})x_{n}\right]
\nonumber\\
&&\times\exp\left[-\frac{i}{2}\sum\limits_{n=1}^{m-1}(\beta_1-\beta_{k_n})
x^2_{n}-i\sum\limits_{n=1}^{m-2}(\beta_1-\beta_{k_n})x_{n}\sum\limits_{j=n+1}^{m-1}x_{j}\right]
\nonumber\\&&
\times2\pi\delta\left[\sum\limits_{n=1}^{m-1}(\beta_1-\beta_{k_n})x_{n}\right]~.
\end{eqnarray}
The subsequent analysis of the multiple integral in
({\ref{dm1-1}}) essentially depends on how much of the indices
$k_n$ are equal unity. At first we will consider the case when all
indices are different from unity. Subsequently the integral with
an arbitrary set of indices will be evaluated. Note that in this
subsection we do not use the condition that the slope $\beta_1$ is
extremal. However in the next subsection this assumption becomes
essential.

\subsection{The case with $k_n\neq 1$ for all $n$}

We carry out a new change of integration variables in such a way
that the argument of the $\delta$-function in (\ref{dm1-1})
depends on a single new variable:
\begin{eqnarray}\label{xy}
y_i&=&\sum_{n=1}^{i}(\beta_1-\beta_{k_n})x_n ~,
\qquad
i=1, \, 2, \, \ldots , \, m-1~.
\end{eqnarray}
The integration limits in the new variables has a simple form due
to the fact that $\beta_1$ has extreme value compared with all
another slopes. For the sake of definiteness we assume that
$\beta_1=\max_j \beta_j$, then
\begin{eqnarray}
y_{m-1}&\in&(0,\infty)~,\nonumber\\
y_{i}&\in&(0,y_{i+1}),
\qquad i= 1, \, 2, \, \ldots, \, m-2~.
\end{eqnarray}
The modulus of the Jacobian for this transformation is
\begin{eqnarray}
|J|=\prod_{n=1}^{m-1}\frac{1}{|\beta_1-\beta_{k_n}|} ~.
\end{eqnarray}
Let us denote the multiple integral in (\ref{dm1-1}) as $I$.
Then in new variables we have
\begin{eqnarray}
I=2\pi|J|\int_{0}^{\infty} dy_{m-1} \, \delta(y_{m-1})
\int_0^{y_{m-1}}dy_{m-2}\cdots
\int_0^{y_2}dy_1~f(y_1,y_2,\ldots,y_{m-2},y_{m-1})~,
\end{eqnarray}
where $f(y_1,y_2,\ldots,y_{m-2},y_{m-1})$ is a regular (smooth)
function of all its arguments. One can see that the integration
over $dy_{m-1}$ with $\delta(y_{m-1})$ in the integrand implies
that $y_{m-1} \rightarrow 0$. This contracts  the integration
range over all other variables to zero. Thus, the entire integral
$I$ is zero.

\subsection{The case with arbitrary set of indices}

Let us assume that ($p-1$) of indices in (\ref{dm1-1}) are equal
to one, where $p\le m$. Taking into account the obvious
restrictions ($k_1\neq 1$ and $k_{m-1}\neq 1$ and $k_{i+1}\ne
k_i$), one obtains a limitation for $p$: $p\le \frac{1}{2}m$ for
even $m$ and $p\le \frac{1}{2}(m-1)$ for odd $m$. In order to
evaluate $I$ in this case we need new notations. Let us introduce
a string of integers ${\cal S}=\{s_1,s_2,\ldots, s_{p-1}\}$ that
includes {\it all}\/ the labels $s_j$ such that $k_{s_j}=1$. It
is an ordered set, so that $s_{i+1} > s_i$. The complementary
string ${\cal C}=\{c_1,c_2,\ldots,c_{m-p}\}$ includes all labels
$c_j$ such that $k_{c_j}\neq 1$ and also is ordered: $c_{i+1} >
c_i$. The multiple integral in (\ref{dm1-1}) is
\begin{eqnarray}\label{second1}
I&=&\int_0^{\infty}dx_{c_1}\int_0^{\infty}dx_{c_2}\cdots
\int_0^{\infty}dx_{c_{m-p}}~
\exp\left[i\sum\limits_{n\in{\cal C}}
(\varepsilon_1-\varepsilon_{k_n})x_{n}\right]
\nonumber\\&&
\times\exp\left[-\frac{i}{2}
\sum\limits_{n\in{\cal C}}(\beta_1-\beta_{k_n})
x^2_{n}-i\sum\limits_{{n\in{\cal C}}\atop{n\ne m-1
}}(\beta_1-\beta_{k_n})x_{n}\sum\limits_{{j>n}\atop{j\in{\cal
C}}}x_{j}\right]
2\pi\delta\left[\sum\limits_{n\in{\cal C}}
(\beta_1-\beta_{k_n})x_{n}\right]
\nonumber\\ &&\times
\int_0^{\infty} dx_{s_1}\int_0^{\infty} dx_{s_2}\cdots\int_0^{\infty}
dx_{s_{p-1}}\exp\left[-i\sum\limits_{{n\in{\cal C}}\atop{n\ne m-1
}}\sum_{{j>n}\atop{j\in {\cal S}}}(\beta_1-\beta_{k_n})x_{n}x_j
\right].
\end{eqnarray}
The integration variables belonging to ${\cal S}$ string enter
exponent linearly (while other variables provide quadratic terms
as well). This allows us to carry out integration in semiinfinite
interval using the formula:
\begin{eqnarray}\label{soh}
\int_{0}^{\infty}e^{ikx} dk=i{\cal P}\frac{1}{x}+\pi\delta(x) ~.
\end{eqnarray}
Here ${\cal P}\frac{1}{x}$ indicates integration in the
principal value sense. After this (\ref{second1}) reduces to
\begin{eqnarray} \label{second}
\nonumber\\
I&=&\int_0^{\infty}dx_{c_1}\int_0^{\infty}dx_{c_2}\cdots
\int_0^{\infty}dx_{c_{m-p}}~\exp\left[i\sum
\limits_{n\in{\cal C}}(\varepsilon_1-\varepsilon_{k_n})x_{n}\right]
\nonumber\\&&
\times\exp\left[-\frac{i}{2}
\sum\limits_{n\in{\cal C}}(\beta_1-\beta_{k_n})
x^2_{n}-i\sum\limits_{{n\in{\cal C}}\atop{n\ne m-1
}}(\beta_1-\beta_{k_n})x_{n}\sum\limits_{{j>n}\atop{j\in{\cal
C}}}x_{j}\right] 2\pi\delta\left[\sum\limits_{n\in{\cal C}}
(\beta_1-\beta_{k_n})x_{n}\right]
\nonumber\\ &&\times
\prod_{j\in{\cal S}}\left[\pi\delta
\left(-\sum \limits^{n<j}_{{n\in{\cal C}\atop{n\ne m-1}}}
(\beta_1-\beta_{k_n})x_{n}\right) +i{\cal
P}\frac{1}{-\sum\limits^{n<j}_{{n\in{\cal C}\atop{n\ne m-1}}}
(\beta_1-\beta_{k_n})x_{n}}
\right] .
\end{eqnarray}

Now a change of variables (\ref{xy}) is conveniently modified to
\begin{eqnarray}\label{xy1}
y_i&=&\sum^{i}_{n=1}(\beta_1-\beta_{k_{c_n}})x_{c_n},
\qquad i=1, \, 2, \ldots, \, m-p ~,
\nonumber\\
y_{m-p}&\in&(0, \, \infty) ~,
\nonumber\\
y_i &\in&(0, \, y_{i+1}) ~,
\qquad i=1, \, 2, \ldots, \, m-p-1 ~.
\end{eqnarray}
The Jacobian modulus is
\begin{eqnarray}\label{Jxy1}
|J|=\prod_{n\in{\cal C}}\frac{1}{|\beta_1-\beta_{k_n}|} ~.
\end{eqnarray}
Each of delta-functions in formula (\ref{second}) depend only
on single new variable $y_i$, so that this formula is cast as
\begin{eqnarray}\label{I}
I&=&2\pi|J|\int_0^{\infty}dy_{m-p}~\delta(y_{m-p})
\int_0^{y_{m-p}}dy_{m-p-1}\cdots\int_0^{y_2}dy_1 \,
f(y_1,y_2,\ldots,y_{m-p})
\nonumber\\
&&\times~\prod_{j=1}^{p-1}\left[\pi\delta(-y_{s_j-j})+i{\cal
P}\frac{1}{-y_{s_j-j}}
\right] ,
\end{eqnarray}
where $f(y_1,y_2,\ldots,y_{m-p})$ is a regular function of all
its arguments. As in previous subsection, the integration over
$dy_{m-p}$ with $\delta$-function contracts to one point, namely
zero, the range of integration over all other variables; thus it
could be said that the contribution from the ${\cal P}$-terms is
zero because of identity
\begin{eqnarray} \label{idP}
\int_0^y {\cal P} \frac{1}{x} \, f(x) \, dx \rightarrow 0
\end{eqnarray}
for $y \rightarrow 0$ and $f(x)$ non-singular at $x=0$. Therefore
the entire integral is different from zero only if integrand is
singular function of all its variable. It could be only if the
number of integrals in (\ref{I}) equals the number of
$\delta$-functions in integrand. This reasoning give us the
condition $m-p=p-1+1$, i.e. $m=2p$. This means that only even
terms in the expansion (\ref{ITSE21}) give non-zero
contributions. The string ${\cal S}$ consists of
$\left(\frac{1}{2}m-1\right)$ numbers. Taking into account the
inequalities $k_{i+1}\neq k_i$, $k_1\neq 1$, $k_{m-1}\neq 1$ we
obtain the necessary condition for indices in (\ref{dm1-1}):
\begin{eqnarray}
k_{2j}=1 \quad {\rm for} \quad j=1, \, 2, \, \ldots,
\, \frac{1}{2} m - 1 ~.
\end{eqnarray}
In other words the following indices have the value 1:
\begin{eqnarray}
k_2, \: k_4, \: k_6, \ldots k_{m-4}, \: k_{m-2} ~.
\end{eqnarray}

\subsection{Summation of non-zero contributions}

For an arbitrary term in (\ref{ITSE21}) we obtain
\begin{eqnarray}
d_1^{2p-1}&=&0~,
\nonumber\\
d_1^{2p}&=&{(-1)}^p2{\pi}^p\sum\limits_{{k_{2p-1}\neq
1}}^NV_{1k_{2p-1}}V_{k_{2p-1}1}\ldots\sum\limits_{k_3\neq 1}^{N}
V_{1k_3}V_{k_31}
\sum\limits_{k_1\neq 1}^N V_{1k_1}V_{k_11}
\nonumber \\&&\times
\prod_{j=1}^{p}\frac{1}{|\beta_1-\beta_{k_{2j-1}}|}
\int_0^{\infty}dy_p\int_{0}^{y_p}dy_{p-1}\cdots\int_0^{y_2}dy_{1}
\nonumber\\ &&\times
f(y_1,y_2,\ldots,y_{p})\prod_{i=1}^{p} \delta(y_{i}) ~,\qquad p=1,
\, 2, \, \ldots .
\end{eqnarray}
The product of $\delta$-functions in the last expression makes
the integrand to be symmetrical function with respect to
arbitrary permutations of the integration variables
$\{y_1,y_2,\ldots,y_p\}$. Besides this, the integrand is an even
function of any of its argument that allows us to extend the
limits of integration:
\begin{eqnarray}
d_1^{2p}&=&{(-1)}^p \, 2{\pi}^p \, \frac{1}{2p\, !} \,
\sum\limits_{{k_{2p-1}\neq 1}}^NV_{1k_{2p-1}}V_{k_{2p-1}1}\ldots
\sum\limits_{k_3\neq 1}^{N} V_{1k_3}V_{k_31}
\sum\limits_{k_1\neq 1}^N
V_{1k_1}V_{k_11}
\nonumber \\&&\times
\prod_{j=1}^{p}\frac{1}{|\beta_1-\beta_{k_{2j-1}}|}
\int_{-\infty}^{\infty}dy_1\int_{-\infty}^{\infty}dy_2\cdots
\int_{-\infty}^{\infty}dy_{p} \:
f(y_1,y_2,\ldots,y_{p})\prod_{i=1}^{p} \delta(y_{i})
\nonumber\\
&=&\frac{{(-\pi)}^p}{p!}\sum\limits_{{k_{2p-1}\neq 1}}^N
V_{1k_{2p-1}}V_{k_{2p-1}1} \ldots
\sum\limits_{k_1\neq 1}^N V_{1k_1}V_{k_11}
\prod_{j=1}^{p}\frac{1}{|\beta_1-\beta_{k_{2j-1}}|}
\nonumber \\
&=&\frac{1}{p \, !}{\left(\sum\limits_{{k\neq 1}}^N\frac{-\pi
V_{1k}V_{k1}}{|\beta_1-\beta_{k}|}\right)}^p~.
\end{eqnarray}
Here we used the property $f(0,0,\ldots,0)=1$.

For the survival amplitude in the limit $n\rightarrow\infty$ we
have the {\it exact}\/ expression:
\begin{eqnarray} \label{survamp}
a_1^{(\infty)}&=&1+\sum\limits_{p=1}^{\infty}\frac{1}{p \, !}
{\left(\sum\limits_{k\neq 1}^{N}\frac{-\pi
V_{1k}V_{k1}}{|\beta_1-\beta_k|}\right)}^p=
\exp\left(-\pi\sum\limits_{k\neq 1}^{N}\frac{ V_{1k}V_{k1}}
{|\beta_1-\beta_k|}\right) .
\end{eqnarray}
Finally, for the survival probability we obtain BE formula:
\begin{eqnarray}\label{GLZ}
P_{11}={|a_1(\infty)|}^2=\exp\left(-2\pi\sum \limits_{k\neq 1}^{N}
\frac{ V_{1k}V_{k1}}{|\beta_1-\beta_k|}\right) .
\end{eqnarray}

\section{Extension of
the approach
to different degenerate cases} \label{degen}

In this section we assume the presence of a special property of a
Hamiltonian compared to general treatment of previous section.
Namely, we presume degeneracy of the potential curves. As above we
consider the situation when the initially populated state 1 has
the largest $(\beta_1=\max_j \beta_j)$ or the smallest
$(\beta_1=\min_j \beta_j)$ of all slopes, except slopes for the
states $1,2\ldots,n$ ($j \neq 1,2,\ldots n$). In other words, we
presume degeneracy of extreme slopes, $\beta_1 =
\beta_2=\cdots=\beta_n $, or, in yet other words, there are
$n$ parallel curves with extreme slope. It is natural also to
presume that the parallel curves are not coupled, i.e. $V_{ij}=0
$ for ($i,j=1,2,\ldots,n$).

The particular case when two bands of parallel potential curves
received some attention in the literature
\cite{band0,band1,band2,band3}.

Subsequently we consider yet more special situation that the
extreme slope curves are not only parallel, but fully degenerate,
i.e. $\varepsilon_1=\varepsilon_2=\cdots=\varepsilon_n$.

\subsection{The case of parallel diabatic potential
curves with extremal slope}

In this subsection we consider the case of $n$ diabatic potential
curves with the same extreme slope $\beta_i=\beta$
($i=1,2,\ldots,n$) and $\beta =
\max_{k>n} \{\beta_k \}$ or $\beta = \min_{k>n} \{
\beta_k \}$. We also assume that $\varepsilon_i\neq\varepsilon_j$ and
$V_{ij}=0$, where ($i\ne j$) and ($i,j=1,2,\dots,n$). Such a
model for $n=2$ was considered in our previous work \cite{No-go},
where transition probablity $P_{12}$ for ($\varepsilon_2 >
\varepsilon_1$) was considered; now we concentrate on the survival
probability. We will prove the formula for survival probability
on the diabatic potential curve with extremal slope for $n=2$.
The proof for arbitrary $n$ might be carry out similarly.

The survival amplitude $a_1^m$ is again given by general formulas
(\ref{ITSE21}) and (\ref{dm1-1}) but the subsequent analysis is a
little  more complicated. The string of integers ${\cal S}$ is
introduced as in the previous section. Besides this, we introduce
a string of integers ${\cal R}=\{r_1,r_2,\ldots,r_g\}$, which
includes all labels such that $k_{r_j}=2$. It is also an ordered
set: $r_{i+1}
> r_i$. The complementary string ${\cal
C}=\{c_1,c_2,\ldots,c_{m-p-g}\}$ includes all labels $c_j$ such
that $k_{c_j}\neq 1,2$ and also is ordered: $c_{i+1} > c_i$. The
dimensions of these strings have to satisfy the conditions:
\begin{eqnarray}\label{ne}
p+g&\le&\frac{1}{2} m \qquad \mbox{for even }m,\nonumber\\
p+g&\le& \frac{1}{2}(m-1) \qquad \mbox{for odd } m~,
\end{eqnarray}
otherwise one or more of the couplings in (\ref{dm1-1}) is zero.
The multiple integral in (\ref{dm1-1}) is in this case after
integration, given through the formula (\ref{soh}):
\begin{eqnarray}
I&=&\int_0^{\infty}dx_{c_1}\int_0^{\infty}dx_{c_2}\cdots
\int_0^{\infty}dx_{c_{m-p-g}}~\exp\left[i\sum\limits_{n\in{\cal C}}
(\varepsilon_1-\varepsilon_{k_n})x_{n}\right]
\nonumber\\&&\times\exp\left[-\frac{i}{2}
\sum\limits_{n\in{\cal C}}(\beta_1-\beta_{k_n})
x^2_{n}-i\sum\limits_{j\in{\cal C}}x_j\sum\limits^{n<j}_{n\in{\cal
C}}(\beta_1-\beta_{k_n})x_{n}\right]
2\pi\delta\left[\sum\limits_{n\in{\cal C}}
(\beta_1-\beta_{k_n})x_{n}\right]
\nonumber\\&&
\nonumber\\ &&\times
\prod_{j\in{\cal
S}}\left[\pi\delta\left(-\sum\limits^{n<j}_{n\in{\cal
C}}(\beta_1-\beta_{k_n})x_{n}\right)+i{\cal
P}\frac{1}{-\sum\limits^{n<j}_{n\in{\cal
C}}(\beta_1-\beta_{k_n})x_{n}} \right]
\nonumber\\
&&\times\prod_{j\in{\cal
R}}\left[\pi\delta\left(-\sum\limits^{n<j}_{n\in{\cal
C}}(\beta_1-\beta_{k_n})x_{n}+(\varepsilon_1-\varepsilon_2)\right)+
i{\cal P}\frac{1}{-\sum\limits^{n<j}_{n\in{\cal C}}
(\beta_1-\beta_{k_n})x_{n}+(\varepsilon_1-\varepsilon_2)}\right]~.
\end{eqnarray}

We realize the change of variables by formula (\ref{xy1}) with
the same Jacobian modulus (\ref{Jxy1}), but now the total amount
of variables is ($m-p-g$). Note that every delta-function after
such transformation depends on only one variable. In new
variables the multiple integral is given by the expression:
\begin{eqnarray}\label{I1}
I&=&2\pi|J|\int_0^{\infty}dy_{m-p-g}~\delta(y_{m-p-g})
\int_0^{y_{m-p-g}}dy_{m-p-g-1}\cdots\int_0^{y_2}dy_1 \,
f(y_1,y_2,\ldots,y_{m-p-g})
\nonumber\\
&&\times~\prod^{p-1}_{j=1}\left[\pi\delta(-y_{s_j-j-\alpha_j})+i{\cal
P}\frac{1}{-y_{s_j-j-\alpha_j}}
\right]
\nonumber\\
&&\times~\prod^{g}_{j=1}\left[\pi\delta(-y_{r_j-j-\beta_j}+
\varepsilon_1-\varepsilon_2)+i{\cal P}\frac{1}{-y_{r_j-j-\beta_j}+
\varepsilon_1-\varepsilon_2} \right]~.
\end{eqnarray}
Here $\alpha_j$ is the number of the elements of the string ${\cal
R}$ which are less than $s_j$, $\beta_j$ is the number of the
elements of string ${\cal S}$ which are less than $r_j$ and
$f(y_1,y_2,\ldots,y_{m-p-g})$ is a regular function of all its
arguments. Note that all $\delta$-functions in the integral
depend on different variables.

The integration over $dy_{m-p-g}$ with $\delta$-function contracts
to one point, namely zero, the range of integration over all
other variables. Thus it could be said that the contribution from
the ${\cal P}$-terms is zero because of identity (\ref{idP}).
Furthermore the contribution from $\delta$-functions in the
second product in (\ref{I1}) is zero. The multiple integral is
different from zero only if integrand is singular function of
every integration variable. This only happens if the number of
integrals in (\ref{I1}) equals the number of $\delta$-functions
in the integrand. This reasoning give us the condition
$m-p-g=p-1+1$, i.e. $m=2p+g$.
Note that if $g\ne0$ this condition contradicts (\ref{ne}). Thus,
this implies that $g=0$. Thereby we come to the same result:
$m=2p$ as in previous section. Besides this, we obtain the
complementary condition $k_j\ne 2$ for $j=1,2,\ldots,m-1$.

The same calculations as in non-degenerate case give us the
survival probability
\begin{eqnarray}\label{GLZ1}
P_{11}={|a_1(\infty)|}^2=\exp\left(-2\pi\sum
\limits_{k\neq 1,2}^{N}
\frac{ V_{1k}V_{k1}}{|\beta_1-\beta_k|}\right) .
\end{eqnarray}
For more  general case of $n$-fold degeneracy $(n < N)$ of
extreme slope potential curves we similarly obtain
\begin{eqnarray}\label{GLZ1n}
P_{jj}={|a_1(\infty)|}^2=\exp\left(-2\pi\sum
\limits_{k\neq 1,2,\ldots,n}^{N}
\frac{ V_{jk}V_{kj}}{|\beta_j-\beta_k|}\right) \qquad j=1,2\ldots,n.
\end{eqnarray}
In case when a band of parallel potential curves is crossed by a
single curve ($n=N-1$)  formula (\ref{GLZ1n}) reproduces an early
result by Demkov and Osherov \cite{Dem-Osh}.

\subsection{The case of merged diabatic potential curves
with extremal slope} \label{merged}

Consider the case when we have $n$ diabatic potential curves with
the same slope $\beta_i=\beta$ ($i=1,2,\ldots,n$) and $\beta =
\max_{k>n} \{\beta_k \}$ or $\beta = \min_{k>n} \{
\beta_k \}$. As distinct from previous subsection we assume that
 $\varepsilon_i=\varepsilon $ ($i=1,2,\ldots,n$).
This means that the potential curves $1,2,\ldots,n$ are merged.
At first we will obtain expressions for survival probabilities for
$n=2$ and then will generalize them for arbitrary $n$.

In the case of two merged diabatic curves with extremal slope we
assume the following conditions for couplings
\begin{eqnarray} \label{merg}
V_{2i}&=&c_2 V_{1i}
\end{eqnarray}
with some $i$-independent constant $c_2$. Acting further as in
non-degenerate case we obtain restrictions for the coefficients:
\begin{eqnarray} \label{prop1}
k_{2j}=1,2 \quad {\rm for} \quad j=1, \, 2, \, \ldots,
\: \frac{1}{2} m - 1 ~.
\end{eqnarray}
For an arbitrary term in (\ref{ITSE21}) we have after integrating:
\begin{eqnarray}\label{merg1}
 d_1^{2p}&=&\frac{{(-\pi)}^p}{p!}
\sum\limits_{k_{2p-1}\neq 1, 2}^NV_{1k_{2p-1}}
\sum\limits_{k_{2p-2}=1}^2V_{k_{2p-1}k_{2p-2}}\ldots
\sum\limits_{k_2=1}^2V_{k_3k_2}
\sum\limits_{k_1\neq 1,2}^N
V_{k_2k_1}V_{k_11} \nonumber \\
&&\times
\prod_{j=1}^p\frac{1}{|\beta_1-\beta_{k_{2j-1}}|}~,\nonumber\\
d_1^{2p-1}&=&0~.
\end{eqnarray}
Due to the property (\ref{prop1}), summations over two terms,
$\sum_1^2$, emerge here. Now we use condition (\ref{merg}) to get:
\begin{eqnarray}
\sum\limits_{k_{2j}=1}^{2}V_{k_{2j-1}k_{2j}}V_{k_{2j}k_{2j+1}}=
V_{k_{2j-1}1}V_{1k_{2j+1}}+V_{k_{2j-1}2}V_{2k_{2j+1}}=
\left(1+c_2^2\right)V_{k_{2j-1}1}V_{1k_{2j+1}} ~.
\end{eqnarray}
Then formula (\ref{merg1}) is rewritten as
\begin{eqnarray}
d_1^{2p}&=&\frac{{(-\pi)}^p}{p!}{(1+c_2^2)}^{p-1}\sum\limits_{k_{2p-1}\neq
1,2}^NV_{1k_{2p-1}}V_{k_{2p-1}1}\ldots\sum\limits_{k_1\neq 1, \:
2}^N V_{1k_1}V_{k_11}\left(\prod_{j=1}^{p}
\frac{1}{|\beta_1-\beta_{k_{2j-1}}|}\right)
\nonumber
\\ &=&\frac{1}{1+c_2^2}{\left(\sum\limits_{k\neq 1, \: 2}^N
\frac{-(1+c_2^2)\pi V_{1k}V_{k1}}{|\beta_1-\beta_k|}\right)}^p\frac{1}{p!} ~,
\nonumber \\
d_1^{2p+1}&=&0 ~.
\end{eqnarray}
Obviously, $d_1^{2p}$ are terms in the expansion of an exponent,
\begin{eqnarray} \label{expsum}
\frac{1}{{(1+c_2^2)}} \exp \left(\sum\limits_{k\neq 1,2}^{N}
\frac{-{(1+c_2^2)}\pi V_{1k}V_{k1}}{|\beta_1-\beta_k|}\right) ~.
\end{eqnarray}
However, the first term in formula (\ref{ITSE21}) is 1, that is
different from the first term in the expansion of expression
(\ref{expsum}). This is easily taken into account. For survival
probability we thus obtain
\begin{eqnarray} \label{sumPP11}
P_{11}&=&\frac{1}{(1+c_2^2)^2}
\left[\exp \left( -(1+c_2^2) \, \sum\limits_{k\neq 1,2}^{N}
\frac{\pi V_{1k}V_{k1}}{|\beta_1-\beta_k|}\right)+c_2^2\right]^2 .
\end{eqnarray}

This result may be easily generalized to the case of $n$-fold
degeneracy of the extreme slope potential curves with an
arbitrary $n$. A simple generalization is possible under
conditions
\begin{eqnarray} \label{Vcond}
 V_{kj}&=&c_kV_{1j} ~, \qquad j>n ~, \qquad k=1,\ldots,n ~,
\end{eqnarray}
which state that the interaction of degenerate states $1, \, 2, \,
\ldots, \, n$ with non-degenerate states ($j>n$) exhibit the same
$j$-pattern, up to common factors $c_k$. Under these conditions
for an arbitrary term in (\ref{ITSE21}) we obtain
\begin{eqnarray}\label{fd1}
d_1^{2p}&=&{(-1)}^p C^{2p-2}\sum\limits_{k_{2p-1}\neq
1,\:2,\ldots,n}^NV_{1k_{2p-1}}V_{k_{2p-1}1}\ldots\sum
\limits_{k_1\neq 1,\:2,\ldots,n}^N
V_{1k_1}V_{k_11}\frac{{\pi}^p}{p!}\left(\prod_{j=1}^{p}
\frac{1}{|\beta_1-\beta_{k_{2j-1}}|}\right)
\nonumber \\
&=&\frac{1}{C^2}{\left(\sum\limits_{k\neq 1, \: 2,\dots,n}^N
\frac{-C^2\pi V_{1k}V_{k1}}{|\beta_1-\beta_k|}\right)}^p\frac{1}{p!} ~,
\nonumber\\
d_1^{2p+1}&=&0 ~,
\end{eqnarray}
where $C^2=\sum_{k=1}^n c_k^2$. For survival probability here we
have
\begin{eqnarray} \label{Pnngen}
P_{11}=C^{-4}
\left[\exp\left( -C^2 \, \sum\limits_{k\neq 1,2,\ldots,n}^{N}
\frac{\pi V_{1k}V_{k1}}{|\beta_1-\beta_k|}\right)
+C^2-1\right]^2 .
\end{eqnarray}

We now turn to evaluation of transition probabilities between
degenerated states $1, \, 2, \, \ldots, \, n$. The expansion
terms $d_1^m$ (\ref{fd1}) in fact do not depend on which of
degenerate states is initially populated. Formally there is
subscript 1 in $d_1^m$ that indicates initial population, but it
could be replaced by any  $j=2, \, 3, \, \ldots, \, n$ without any
other change in formulas, except for changing couplings
$V_{1k_{2p-1}}$ to $V_{j{k_{2p-1}}}$~.

However, there is difference in the first term of the
perturbative expansion (\ref{ITSE21}) that explicitly indicates
the initial population. Taking this into account, it is easy to
write down expression for probabilities of transitions within the
submanifold of degenerate states:
\begin{eqnarray} \label{Pintraband}
P_{1j}=\frac{c_j^2}{C^4}{\left[\exp\left(-C^2
\sum\limits_{k \neq 1, 2,\ldots,n}^{N}
\frac{ \pi V_{1k}V_{k1}}{|\beta_1-\beta_k|}\right)-1\right]}^2 ,
\qquad  j = 2, \ldots, \, n ~.
\end{eqnarray}

\subsection{Alternative derivation via orthogonalization}

Now we consider alternative scheme of derivation for the case when
we have only two diabatic potential curves with the same slope
$\beta_1=\beta_2=\beta$ and $\beta = \max_k \{\beta_k \}$ or
$\beta = \min_k \{ \beta_k \}$, and
$\varepsilon_1=\varepsilon_2$. As the conditions on couplings we
again use formula (\ref{merg}).

Introduce a new basis with the states $| \tilde{1} \rangle$ and $|
\tilde{2} \rangle$:
\begin{eqnarray} \label{concoup2}
| \tilde{1} \rangle = h \left( c_2 | 1 \rangle -  | 2 \rangle
\right) ,
\\
| \tilde{2} \rangle = h \left( | 1 \rangle + c_2 | 2 \rangle
\right) ,
\\
h = \left(1 + c_2^2 \right)^{-1/2}
\end{eqnarray}
instead of states $|1 \rangle$ and $|2 \rangle$; all other states
coincide in the new and old bases. Obviously, the new basis is
orthonormal. The non-diagonal elements of Hamiltonian matrix with
the states $| \tilde{1} \rangle$ are all zero:
\begin{eqnarray} \label{Hzero}
\langle \tilde{1} | H | j \rangle = 0 ~,
\qquad j = \tilde{2}, \: 3, \: 4, \: \ldots \: N ~;
\end{eqnarray}
in other words state vector $|\tilde{1} \rangle$ is orthogonal to
all vectors $H | j \rangle$. This means that the state
$|\tilde{1}\rangle$ is fully decoupled from all the other states.
The diagonal elements of Hamiltonian matrix remain the same in
new basis. In terms of S matrix this could be written as
\begin{eqnarray} \label{Szero}
\langle \tilde{1} | S | \tilde{1} \rangle & = & 1 ~,
\\
\langle \tilde{2} | S | \tilde{2} \rangle & = &
\exp\left(-\pi\sum \limits_{k\neq 1,2}^{N}
\left|\langle \tilde{2} | H | k \rangle \right|^2
\frac{1}{|\beta-\beta_k|}\right) = {\cal D} ~,
\end{eqnarray}
where we define
\begin{eqnarray} {\cal D} &\equiv&
\exp\left(-\pi \left(1+c_2^2 \right)  \sum \limits_{k\neq 1,2}^{N}
\frac{{|V_{1k}|}^2}{|\beta-\beta_k|}\right)~.
\end{eqnarray}
Here we used the result (\ref{survamp}) obtained above for the
non-degenerate case. The desired S~matrix element in the original
basis is
\begin{eqnarray}
\langle 1 | S | 1 \rangle & = & h^2 c_2^2
\langle \tilde{1} | S | \tilde{1} \rangle +
h^2 \langle \tilde{2} | S | \tilde{2} \rangle = h^2 (c_2^2 +
{\cal D}) ~.
\end{eqnarray}
This gives the state-to-state transition probability
\begin{eqnarray} \label{ortPP11}
P_{11} &=& h^4 (c_2^2 +  {\cal D})^2 ~,
\end{eqnarray}
which coincides with the earlier obtained result in
(\ref{sumPP11})~.

\subsection{Fully degenerate multistate model} \label{fully}

Consider the case when two fully degenerated bunches of
potential curves cross each other. The Hamiltonian of
this model has form:
\begin{eqnarray}
H=
 \left(
\begin{array}{ccccccc}
E_1&0&\cdots&0&V&\cdots&V\\
\vdots&&\ddots &\vdots&\vdots&&\vdots\\
0&0&\cdots&E_1&V&\cdots&V\\
V&V&\cdots&V&E_2&\cdots&0\\
\vdots&\vdots&&\vdots&\vdots&\ddots&\vdots\\
V&V&\cdots&V&0&\cdots&E_2\\
\end{array}
\right)
\end{eqnarray}
Let $n$ be the number of potential curves with energy $E_1 =
\beta_1 t$ and $m$ be the number of potential curves with energy
$E_2 = \beta_2 t$. The Hamiltonian matrix has dimension $(n+m)
\times (n+m)$. Some transition probabilities for this model can
be written down straight off as particular cases of formulas
(\ref{Pnngen}) and (\ref{Pintraband}). The survival probabilities
are
\begin{eqnarray}\label{NNMM}
P_{jj}&=&\frac{1}{n^2}{\left(p^{nm/2}+n-1\right)}^2
\qquad
j=1,\ldots,n ~,
\nonumber \\
P_{jj}&=&\frac{1}{m^2}{\left(p^{nm/2}+m-1\right)}^2
\qquad
j=n+1,\ldots,n+m ~.
\end{eqnarray}
The intraband transition probabilities are
\begin{eqnarray}\label{NKMK}
P_{jk}&=&\frac{1}{n^2}{\left(p^{nm/2}-1\right)}^2
\qquad
j\neq k, \qquad j,k=1,\ldots, n ~,
\nonumber \\
P_{jk}&=&\frac{1}{m^2}{\left(p^{nm/2}-1\right)}^2\qquad j\neq
k,\qquad j,k=n+1,\ldots,n+m ~,
\end{eqnarray}
where $p$ is the standard  Landau-Zener probability:
\begin{eqnarray}
p=\exp\left(\frac{-2\pi {|V|}^2}{|\beta_1-\beta_2|}\right) .
\end{eqnarray}

The remaining (interband) probabilities one can obtain by using
the normalization condition:
\begin{eqnarray} \label{norm}
\sum\limits_{j=1}^{n+m}P_{jk}=\sum\limits_{k=1}^{n+m}P_{jk}=1~.
\end{eqnarray}

From general considerations it can be concluded that all
interband transition probabilities are equal, i.e.:
\begin{eqnarray} \label{same}
P_{jk}=P_{jk'} ~, \qquad j=1, \: \ldots, \: n ~,
\qquad
k,k'=n+1, \: \ldots, \: n+m ~,
\nonumber\\
P_{jk}=P_{jk'} ~, \qquad j=n+1, \: \ldots, \: n+m ~,
\qquad
k, \: k'=1, \: \ldots, \: n ~.
\end{eqnarray}

Using (\ref{norm}) and (\ref{same}) we obtain:
\begin{eqnarray}\label{NM}
P_{jk}=P_{kj}=\frac{1}{n m}\left(1-p^{nm}\right) ~,
\qquad
j=1, \: \ldots, \: n ~, \qquad k=n+1, \: \ldots, \: n+m ~.
\end{eqnarray}

Thus in this highly degenerate multistate model there are only
five different state-to-state transition probabilities defined by
expressions (\ref{NNMM}), (\ref{NKMK}) and (\ref{NM}) . This
conclusion as well as quantitative results were tested by
numerical calculations.

\section{Conclusion}

In this paper we consider calculation of state-to-state
transition probabilities in the generalized multistate
Landau-Zener model by summation of perturbation theory series. Due
to specifics of generalized Landau-Zener Hamiltonian (linear
growth with time), some of the integrals emerging in the
pertubative expansions are singular and require special analysis.
The singularities of these integrals are 'useful' in the sense
that they effectively cancel other integrations, such that the
analytical expressions are obtained for each term in the
perturbative expansion. Subsequently, entire infinite series is
summed with the result obtained in closed form. The technique of
such calculations is one of the principal results of the present
study.

The other group of results refers to the degenerate cases. In the
general non-degenerate case we are able to evaluate only two
transition probabilities: the survival probabilities for diabatic
potential curves with maximum and minimum slope. Such a situation
when some state-to-state transition probabilities are expressed
by simple analytical formulas, while others remain unknown is
quite unconventional, although now we know another similar
example: the multistate Coulomb model \cite{Ocoul}. As long as the
degeneracy conditions are introduced, the analytical expressions
for some new state-to-state transition probabilities are obtained.
For the case of extreme degeneracy, when two fully degenerate
bands of diabatic potential curves cross each other, the full set
of state-to-state transition probabilities was derived. Various
degenerate cases are met in practice, for example, in the
treatment of second order effects in Rydberg H atom in
perpendicular electric and magnetic fields \cite{sec}.

\acknowledgements
While this report was almost completed, my co-author, friend and
scientific advisor Valentin N.~Ostrovsky prematurely deceased. I,
(M.V.), would like to express here my deep gratitude to him for
all that he has done for me.

I, (M.V.), am grateful to the Swedish Institute for support.


\begin{thebibliography}{93}

\bibitem{Land}
L.~D.~Landau,  Phys. Z. Sowjetunion {\bf 2}, 46 (1932).

\bibitem{Zen}
C.~Zener,  Proc. R. Soc. Lond. A {\bf 137}, 696 (1932).

\bibitem{M}
E.~Majorana, Nuovo Cimento {\bf 9}, 43 (1932).

\bibitem{St}
E.~C.~G.~St\"{u}kelberg, Helvetica Physica Acta {\bf 5}, 369 (1932).

\bibitem{Dem-Osh}
Yu.~N.~Demkov and V.~I.~Osherov,  Zh. Exp. Teor. Fiz.\/ {\bf 53},
1589 (1967) [ Sov. Phys. - JETP\/ {\bf 26}, 915 (1968)].

\bibitem{Ostr-Nak}
V.~N.~Ostrovsky and H.~Nakamura, Phys. Rev. A {\bf 58}, 4293
(1995).

\bibitem{DOgen1}
Yu.~N.~Demkov and V.~N.~Ostrovsky, Phys. Rev. A {\bf 61}, 032705 (2000).

\bibitem{DOgen2}
Yu.~N.~Demkov and V.~N.~Ostrovsky, J. Phys. B {\bf 34}, 2419 (2001).

\bibitem{BE}
S.~Brundobler and V.~Elser, J. Phys. A {\bf 26}, 1211 (1993).

\bibitem{Shyt}
A.~V.~Shytov, Phys. Rev. A {\bf 70}, 052708 (2004).

\bibitem{SurPr}
M.~V.~Volkov and V.~N.~Ostrovsky, J. Phys. B {\bf 37}, 4069
(2004).

\bibitem{Comment}
B.~E.~Dobrescu and N.~A.~Sinitsyn. J. Phys. B {\bf 39}, 1253 (2006).

\bibitem{reply}
M.~V.~Volkov and V.~N.~Ostrovsky, J. Phys. B {\bf 39}, 1661
(2006).

\bibitem{No-go}
M.~V.~Volkov and V.~N.~Ostrovsky, J. Phys. B {\bf 38}, 907 (2005)

\bibitem{Kay}
Y.~Kayanuma, J. Phys. Soc. Japan {\bf 53}, 108 (1984); {\bf 53},
118 (1983).

\bibitem{Kay1}
Y.~Kayanuma, Phys. Rev. Lett. {\bf 58}, 2934 (1987).


\bibitem{band0}
Yu.~N.~Demkov and V.~N.~Ostrovsky, J. Phys. B {\bf 28} 403 (1995)

\bibitem{band1}
Yu.~N.~Demkov, P.~B.~Kurasov and V.~N.~Ostrovsky, J. Phys. A {\bf
28} 4361 (1995)

\bibitem{band2}
V.~N.~Ostrovsky and H.~Nakamura, Phys. Rev. A {\bf 58} 4293 (1998)

\bibitem{band3}
T.~Usuki Phys. Rev. B {\bf 56} 13360 (1997)

\bibitem{sec}
 V.~N.~Ostrovsky, J. Phys. B {\bf 38}, 1483 (2005).

\bibitem{Ocoul}
V.~N.~Ostrovsky, Phys. Rev. A {\bf 68}, 012710 (1-7) (2003).


\end{thebibliography}
\end{document}